\documentstyle[prl,aps,epsf,psfig,multicol]{revtex}
\begin{document}
\draft 
\title{Proof of the thermodynamical stability of
the E$'$ center in SiO$_2$}%, a 45-year-old puzzle of defects physics}  
\author{Carlo Maria Carbonaro, Vincenzo  Fiorentini, and
Fabio Bernardini}
\address{INFM and Dipartimento di  Fisica, Universit\`a di Cagliari,
Cittadella  Universitaria, I-09042 Monserrato (CA), Italy}
\date{\today} 
\maketitle

\begin{abstract}
The E$'$ center is a paradigmatic radiation-induced defect in SiO$_2$
whose  peculiar EPR and hyperfine activity has been known since over
40 years. This center  has been traditionally identified with a
distorted, positively-charged oxygen vacancy V$_{\rm O}^+$.  However,
no direct proof of the stability of this defect has  ever been
provided, so that its identification is still strongly
incomplete. Here  we prove directly that distorted V$_{\rm O}^+$ is 
metastable and that it satisfies the key requirements for its 
identification as E$'$, such as thermal and optical response, and
activation-deactivation  mechanisms.\end{abstract} 
  % point def.

\begin{multicols}{2}
Understanding defects in solids is a key factor in improving device
performance and materials quality. Defect identification combines
experimental observation and theoretical predictions, and a major
ingredient in this process is the theoretical justification of the
thermodynamical stability of the defect. If this information is
missing, the identification is arguably incomplete or
uncertain. Surprisingly, this is the case for the E$'$ center \cite{1,2}, a
radiation-induced defect first observed experimentally as early as 45
years ago \cite{1} in SiO$_2$, a material of prime current importance in
microelectronics and fiber optics \cite{3}. E$'$ is traditionally identified
with a positively-charged distorted oxygen vacancy V$_{\rm O}^+$
\cite{4},
with important support from calculations of hyperfine couplings
\cite{5} and optical 
activity \cite{6}; its thermodynamical stability, however, was never
theoretically proven, and the mechanisms involved in its activation
and deactivation are still debated. Here, using the ab initio theory
of defect formation in solids \cite{7}, we show that the conditions for E$'$
stability are naturally realized in stoichiometric or
neutron-irradiated SiO$_2$, and conclusively put on firm ground the
identification of the E$'$ center.   

Our model of the stability of E$'$ is based on two native defects: the
oxygen vacancy V$_{\rm O}$ and the oxygen interstitial O$_i$. The motivation is
that E$'$ is observed chiefly (though not only) in neutron-irradiated
material \cite{1,2,3}, where V$_{\rm O}$ and O$_i$ are both abundant. Indeed, we find
that it is their concurrent presence that produces the conditions for
the existence of E$'$, in neutron-irradiated as well as
non-neutron-irradiated material. In the former, vacancies 
V$_{\rm O}$ and interstitials O$_i$ 
are essentially  produced in pairs by knock-on, kick-out events;
 in the latter, they form in thermal equilibrium and, as it turns out,
in similar concentrations.    As will become apparent below, our
argument on E$'$  applies to both cases without modifications, except
for the fact  that neutron-produced defects appear in concentrations
determined by the irradiation dose, whereas the concentration of
thermally formed defects depends on formation energies, which   can be
directly predicted. 

 At a growth temperature T$_{\rm g}$ and with N$_s$ available 
sites, the   equilibrium concentration of a defect is D$= {\rm N}_s \exp 
(-{\rm F}_{\rm form}/{\rm k}_{\rm B} {\rm T}_{\rm g})$. The
formation free energy F$_{\rm form}$ = E$_{\rm form}$ -- T S$_{\rm
form}$ depends \cite{7} on the 
chemical potentials of atoms added or removed, on the defect charge
state, i.e. the charge  released to or captured from the thermodynamic
reservor constituted by the surrounding crystal, and on the
electronic chemical potential $\mu_e$ of the latter. Once the formation
energies of the all relevant defects (vacancy and interstitial in
our case) are known, the defect concentrations
and the  chemical potential $\mu_e$ are
determined self-consistently, subject to charge neutrality, as
detailed in \cite{7}. A specific defect configuration or charge state is
predicted to exist if its formation energy is lower than that of all
other defect states for some value of $\mu_e$. Also, the defect is
metastable if 
a non-zero energetic barrier prevents its deactivation or
disappearance into other lower-energy configurations of the same
defect, or recombination with other defects. The formation energy for
our defects in charge state $Q$ reads    
\begin{equation}
{\rm E}_{\rm form} (Q) = {\rm E}_{\rm tot}^{\rm def} (Q) - 
 {\rm E}_{\rm tot}^{\rm undef}  + Q \mu_e + {\rm M} (Q) + {\rm P}
\end{equation}
where  E$_{\rm tot}^{\rm def}$ and
 E$_{\rm tot}^{\rm undef}$ are the total energies of the defected
and undefected system, respectively, $\mu_e$ is the electron chemical
potential (equaling the Fermi level E$_{\rm F}$ in our T=0 
calculations), M$ (Q)$ is the defect-dependent multipole correction for
 charge state $Q$ of Ref. \cite{8}, P = $\mu_{\rm O}$ for 
V$_{\rm O}$ and P = --$\mu_{\rm O}$ for O$_i$, and $\mu_{\rm O}$ 
is the oxygen
chemical potential. The latter is fixed to stoichiometric conditions,
i.e. at the center of its variation range $\mu_{\rm mol}/2 +
 \Delta {\rm H}/2 < \mu_{\rm O} < \mu_{\rm mol}/2$
determined by the total energy $\mu_{\rm mol}$ of the O$_2$ molecule,
 and the calculated formation enthalpy 
$\Delta {\rm H}$ of  SiO$_2$. Ionization  energies, 
i.e. the energy needed to promote (say) an electron from the valence
band to an empty acceptor level, are defined via total energy
differences of different charge states. Formation entropies are beyond
the scope of the method used here; plausible estimates are used when
needed. 

Energies and forces are accurately calculated from
first-principles within density-functional theory in the local
approximation, using the ultrasoft pseudopotential plane-wave method
as implemented in VASP \cite{9}. An isolated defect is simulated in
periodic boundary conditions via the repeated supercell approach. We
use crystalline $\alpha$-quartz SiO$_2$ supercells of tetragonal symmetry,
comprising 71 to 73 atoms and of linear dimensions 18.49, 16.02, and
20.44 atomic units (theoretical lattice parameters \cite{10}, matching
experiment to about 0.5\%). Atomic geometries of the defects are
optimized for each $Q$ (obtained by removing or adding electrons as
appropriate) until all residual force components in the system are
below 0.02 eV/\AA. No symmetry restriction is imposed on geometry 
optimization. Improving slightly on the setting of Ref.\cite{5}, a (222)
mesh is used \cite{9} for  k-space summation (4 special points in the
supercell  Brillouin zone). Our use of a crystalline SiO$_2$-based model
of the E$'$ defect, which is observed both in amorphous and crystalline
phases, is justified by the closely similar behavior of
several E$'$ variants in crystalline and amorphous SiO$_2$ in experiment
\cite{2,11} as well as theory \cite{4,5}. In addition, the simulated structure
of amorphous silica \cite{new} deviates moderately from that of crystalline
a-quartz SiO$_2$. 

\begin{figure}
\epsfclipon
\epsfxsize=8cm
\epsffile{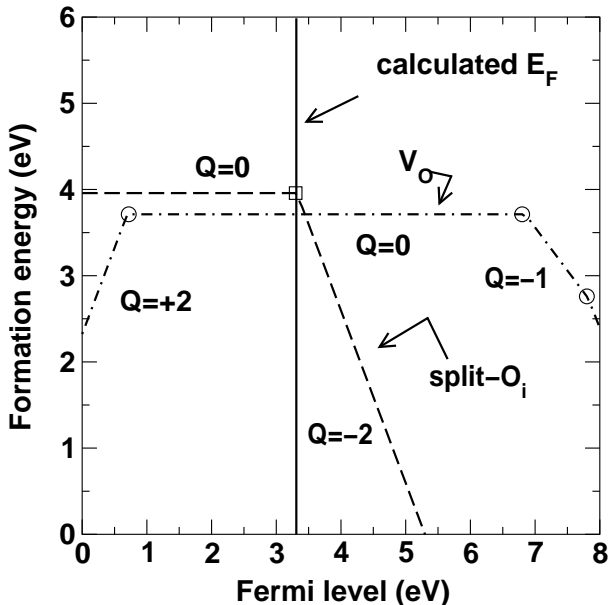}
\caption {Minimum formation energies (eV) of the oxygen vacancy
(dash-dotted line) and split-interstitial oxygen (dashed) as a
function of the Fermi level, and calculated Fermi levels position
(vertical solid). Fermi level zero is the valence band top.}
\label{figura.1}
\end{figure}

Fig.\ref{figura.1} shows the ground state formation energies of the
interstitial O$_i$ 
and the vacancy V$_{\rm O}$. The vacancy acts as a double acceptor (double
donor) in extreme $n$ ($p$) doping conditions, but it is in fact neutral
for most of the Fermi level range. 
 As to the interstitial, we find that an
oxygen atom initially placed near the center of the 
hexagonal channels of quartz, relaxes sideways towards the helical chains
and, after overcoming a small ($\simeq$ 0.2 eV) barrier, it stabilizes
into a split-interstitial (s-O$_i$) configuration with the nearest
bridging oxygen in the helical tetrahedra chain, with $\simeq$ 1 eV of
energy gain with respect to the starting site.  Details will be
discussed elsewhere: here we note the main consequences of this
result:  
{\it a)} the stabilization in the s-O$_i$ configuration (which is
found to be  quite close to that suggested in Ref.\cite{12})  prevents
O$_i$ from easily recombining with vacancies, because the 
detachment from the split-interstitial configuration costs about 1.2
eV; {\it b)} as seen in Fig.\ref{figura.1},  s-O$_i$ is a (negative-$U$)
deep double acceptor with second ionization energy at  3.3 eV above
the valence band top Ev. The Fermi level, calculated  as in \cite{7},
is pinned at E$_{\rm F}$ = E$_v$ + 3.3 eV (vertical solid line in
Fig.\ref{figura.1}). Thus a consequence of vacancy-interstitial pair
formation,  is that moderately $p$-type conditions are
achieved; this is indeed often observed in irradiated samples
\cite{2,11}. In the absence of s-O$_i$, the Fermi level would be at
midgap, E$_v$ + 4.4 eV. From the formation energies one can estimate
the chemical concentrations of s-O$_i$ and V$_{\rm O}$: for a typical
T$_{\rm g}$  of  1500 K, and assuming a  reasonable formation entropy
S$_{\rm form}$= 5 k$_{\rm B}$,  the concentration of both defects is
$\sim$10$^{14}$ cm$^{-3}$.   
This figure matches well E$'$ concentrations measured \cite{11} in
non-neutron irradiated samples after UV, $\gamma$, or X
illumination fall, and therefore  supports the hypothesis that the
vacancy is the parent defect of E$'$. Neutron irradiation of course
produces dose-dependent  \cite{2}, typically much higher
concentrations ($>10^{19}$ cm$^{-3}$). The key point is that, since
the concentrations of vacancies and interstitial are essentially the
same in both cases, and because  only the interstitial has  electrical
activity (via its double acceptor level),  the Fermi level is pinned
at the same value in  both  cases, so that the theory  of E$'$
stability and activation discussed below applies identically to both
cases.  

	It appears from Fig.\ref{figura.1} that the +1 charge state of
the vacancy, 
V$_{\rm O}^+$, is not among the thermodynamically stable ground states
of the defect. Therefore, if this state of V$_{\rm O}$ is to be
identified with E$'$, it must at least be proven metastable; if it is
metastable, a mechanism 
for its creation starting from the ground state (the neutral vacancy)
must be identified. 
As to the first point, since the
+1 vacancy has a formation energy that increases linearly with the
Fermi level while the neutral one remains constant, V$_{\rm O}^+$ 
or a distorted variant thereof, may
only be metastable in a limited range of that variable: we show below
that the Fermi level pinning due to split-O$_i$ produces naturally the
conditions for the metastable existence of E$'$. As to the second
point, experiments indicate that E$'$ is activated by ionizing
radiation \cite{2,3,11}  such as $\gamma$, X, or UV photons shone onto
vacancy-containing samples, or  concurrently with neutron irradiation
(causing knock-on vacancy  creation). Indeed, since the  vacancy
ground state is the neutral  undistorted configuration, the distorted
+1 state (alias E$'$) can  only be accessed by ionization of V$_{\rm
O}^0$: our picture provides consistently such activation mechanism. 

We proceed to study the behavior of the vacancy when subjected to
the undistorted-to-puckered transition as proposed in earlier studies
\cite{4,5,6}. In accord with the results discussed above, we fix the Fermi
level at E$_v$ + 3.3 eV. The creation of a vacancy starting from the
perfect lattice results in  moderate local distortions in both the neutral
and +1 charge states. The puckered configuration is obtained by moving
one of the two vacancy-adjacent Si$_1$ and Si$_2$ atoms (specifically the
``long-bond \cite{4,5} Si$_2$) away from the vacant site, and pushing it
across the basal plane of the incomplete tetrahedron centered on Si2
itself. When Si$_2$ pokes through this triangular constriction, it gets
strongly and suddenly bound to a {\it backbonding} oxygen, O$_b$. Upon
completion of the distortion, Si$_2$ regains 4-fold coordination, and
backbonding O$_b$ becomes 3-fold coordinated (see also \cite{4,5,6}), while it
was originally 2-fold coordinated as all tetrahedron-bridging oxygens
in SiO$_2$. Si$_1$ remains instead 3-fold coordinated: in the +1 charge
state, its dangling bond is half-filled, and causes the observed \cite{1,2}
and predicted \cite{5} EPR signature which identifies E$'$ \cite{2}. 

It may appear at first sight that the puckering distortion should be
symmetric in Si$_1$ and Si$_2$. This is not the case, however, because
of the intrinsic asymmetry of the quartz structure. As already noted
earlier on \cite{4}, only Si$_2$ finds the backbonding oxygen O$_b$ in
the correct position. This applies largely also to amorphous silica,
whose structure is moderately different at the local level from that
of quartz \cite{new}.  The backbonding oxygen is therefore the main 
 stabilizing agent of the E$'$ defect.

\begin{figure}[h]
\epsfclipon
\epsfxsize=8cm
\epsffile{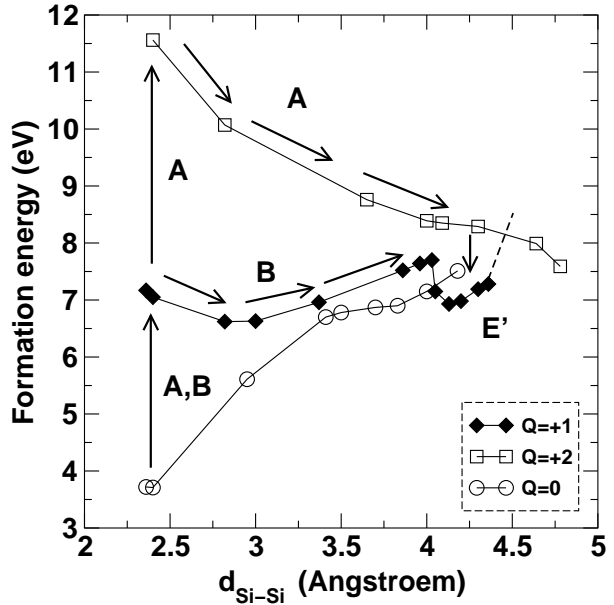}
\label{figura.2}
\caption {Total energy of neutral, positive (E$'$), and doubly positive oxygen
vacancies as a function of the puckering distortion, at the calculated
Fermi level value.} 
\end{figure}

The total
energy of the system in charge state $Q$ is calculated as a function of
the separation between Si$_1$ and Si$_2$. Only the modulus d$_{\rm
Si-Si}$ of the 
Si$_1$-Si$_2$ connecting vector is constrained,  and all other degrees of
freedom are fully relaxed: the minimum energy path is thus mapped out
for the undistorted-puckered transformation in the constrained-d$_{\rm
Si-Si}$
configurational subspace. In Fig.\ref{figura.2} we display the full
level diagram  
for the neutral, +1, and +2 charge states of the vacancy as a function
of the puckering distortion, quantified by
d$_{\rm Si-Si}$. All energy curves
depend on the Fermi level through Eq.1; they can be directly compared
on the same energy scale because E$_{\rm F}$ is fixed at the natural value of
E$_v$ + 3.3 eV determined above.  

The outstanding feature of Fig.\ref{figura.2} is that for the natural
Fermi level of stoichiometric or neutron-irradiated silica, the
candidate E$'$, i.e. distorted V$_{\rm O}^+$, is indeed the stable
defect state for the distorted geometry. We stress that the Fermi level
position is 
essential here: if E$_{\rm F}$ were at midgap, the +1 curve would be 1.1 eV
above its position in Fig.\ref{figura.2}. Then  E$'$ would 
be unstable towards magnetically-inactive V$_{\rm O}^0$. Globally,
 E$'$ is metastable with a confining barrier of 0.8
eV. The  barrier to enter the metastable state is 1.1 eV, and the
undistorted state is lower than the distorted  by 0.3 eV [this
difference to Ref. \cite{5}, where the  distorted state was lower by
the same amount, is possibly due  to our improved
k-sampling]. Clearly, in the absence of   excitation, V$_{\rm O}^+$
will remain trapped in the metastable E$'$ state and will show EPR
activity. When thermally activated to overcome the 
barrier, the puckered center will transform into undistorted, and by
electron capture it will become neutral. Therewith, E$'$ disappears
permanently, and  so does its magnetic activity, because
Si1 and Si2 combine their dangling bonds to bind into a dimer
\cite{4,5,6,10}. (The same deactivation route is not readily available
for the level ordering of Ref. \cite{5}, which implies that {\it i)} E$'$
remains activated at equilibrium, and that {\it ii)}
concurrent barrier jump and electron capture are required to quench
it.)

The calculated puckered-to-undistorted barrier is now compared with an
estimate extracted from the measured relative drop in E$'$ population
upon  isocronal thermal annealing \cite{11} in irradiated silica. In the
simplest model, the distorted-state  population N$_0$ diminishes by a
factor of N/N$_0$ = exp (--R(T) $\tau$) upon annealing over a time
$\tau$  at temperature T, with escape rate R = $\nu$ exp
(-$\Delta$F/k$_{\rm B}$T),  with $\Delta$F the free energy barrier for
escape from the 
puckered state, and $\nu$ an average vibrational frequency in that
state. Using the data \cite{11} for E$'_{\gamma}$ and assuming $\nu
\sim 50$ THz, we   obtain an experimental $\Delta$F$\sim$1.1 eV, in
reasonable agreement with our calculated  $\Delta$F$\sim\Delta$E=0.8
eV; account for the transition entropy, which we neglect, should
further improve agreement since the transition state (through the
tetrahedron basal plane) is severely constrained geometrically, and
has a higher average vibrational frequency.    

Let us now come to E$'$ activation. In the present picture, E$'$ is
created  transforming undistorted V$_{\rm O}^0$ into distorted
V$_{\rm O}^+$ via two routes. The first proceeds on path A in
Fig.\ref{figura.2} with two successive one-photon ionizations of
V$_{\rm O}^0$ into  V$_{\rm O}^{+2}$, followed by non-radiative decay
into E$'$. This path is efficient since the +1 undistorted state is
kept populated by sustained illumination (a much less efficient
double-photon excitation of V$_{\rm O}^0$ into V$_{\rm O}^{+2}$ may
also occur). The  excitation energies for path A are both near 4--4.5
 eV if the ionized electron is transfered to the Fermi level, i.e. to
the E$_{\rm F}$-pinning impurity; if it is promoted to the conduction
band, the energies are instead about 7 eV. Both processes are
possible with X or $\gamma$ radiation, whose energy vastly exceeds
that needed in the transition. In UV irradiation, the center is often
activated by pumping at 5 eV, and clearly only Fermi-level capture
matches this figure. (There are, however, qualitative indications that
the 5 eV excitation may activate E$'$ via alternate routes involving
other pre-existing defects.)

The second excitation route, path B, involves an optical excitation of
V$_{\rm O}^0$ into undistorted V$_{\rm O}^+$, and a thermal excitation
of the latter into the puckered state. The energy difference (0.3 eV)
between the two V$_{\rm O}^+$  states implies that only a fraction of
10$^{-5}$ of the vacancies gets promoted into the distorted state in
equilibrium at room temperature, on sustained illumination. Therefore,
though admissible, this path is preempted by path A. With the level
ordering of Ref. \cite{5}, also path B competes with path
A. (Our ordering, however, matches better the thermal behavior, as
discussed above.)  

Other calculated observables of our E$'$ model are consistent with
previous studies \cite{4,5,6,13}. For instance, the optical absorption
of the neutral undistorted state at 6.9 eV correlates well with the
7.6 eV absorption band usually attributed to the neutral vacancy
\cite{6}. For E$'$ (metastable puckered V$_{\rm O}^+$) we find an
absorption at 4.7 eV (defect-to-conduction promotion) followed by slow
non-radiative decay back into E$'$. Since it is not followed by any
emission or E$'$ deactivation, this absorption correlates with the
broad 5.8-eV band typical of E$'$ \cite{11,13}, which exhibits the
same behavior in experiment. 

In summary, we conclusively put on firm ground the identification of
the singly positive O vacancy in SiO$_2$ with the E$'$ center proving its
thermodynamical stability via first principles calculations.  Our
picture
provides naturally activation and deactivation mechanisms, and other
optical signatures, in agreement with known experimental observations.
In addition, our picture naturally explains the moderate $p$
conditions produced  by irradiation in SiO$_2$.

We acknowledge support from 
Istituto Nazionale per la Fisica della Materia under 
 ``Iniziativa Trasversale Calcolo Parallelo''.

\end{multicols}
\end{document}